# Collaborative human-AI trust (CHAI-T): A process framework for active management of trust in human-AI collaboration

Melanie J. McGrath* [a], Andreas Duenser[a], Justine Lacey[b] & Cécile Paris[a]

[a] CSIRO, Private Bag 10, Clayton South, Victoria, 3169, Australia

[b] CSIRO, GPO Box 2583, Brisbane, Queensland, 4001, Australia

* Corresponding author: Melanie McGrath

Email: melanie.mcgrath@data61.csiro.au

**Author statement**

**Melanie McGrath**: Conceptualization, Methodology, Writing – Original draft, Writing – Review and Editing. **Andreas Duenser**: Conceptualization, Writing – Review and Editing, Project administration. **Justine Lacey**: Writing – Review and Editing, Supervision, Funding acquisition. **Cécile Paris**: Writing – Review and Editing, Supervision, Funding acquisition.

This research did not receive any specific grant from funding agencies in the public, commercial, or not-for-profit sectors.

**Abstract**

Collaborative human-AI (HAI) teaming combines the unique skills and capabilities of humans and machines in sustained teaming interactions leveraging the strengths of each. In tasks involving regular exposure to novelty and uncertainty, collaboration between adaptive, creative humans and powerful, precise artificial intelligence (AI) promises new solutions and efficiencies. User trust is essential to creating and maintaining these collaborative relationships. Established models of trust in traditional forms of AI typically recognize the contribution of three primary categories of trust antecedents: characteristics of the human user, characteristics of the technology, and environmental factors. The emergence of HAI teams, however, requires an understanding of human trust that accounts for the specificity of task contexts and goals, integrates processes of interaction, and captures how trust evolves in a teaming environment over time. Drawing on both the psychological and computer science literature, the process framework of trust in collaborative HAI teams (CHAI-T) presented in this paper adopts the tripartite structure of antecedents established by earlier models, while incorporating team processes and performance phases to capture the dynamism inherent to trust in teaming contexts. These features enable active management of trust in collaborative AI systems, with practical implications for the design and deployment of collaborative HAI teams.

**Keywords**

trust; artificial intelligence; collaborative intelligence; human-AI teaming; trust calibration; team processes

# 1. Introduction

Artificial intelligence (AI) is not always that intelligent. Machines can work well on their own where the problem space is clearly defined, and many tasks can benefit from being automated. However, some tasks require not only the physical strength or powerful processing of AI, but uniquely human capabilities like moral reasoning, flexibility or critical thinking. For example, AI is typically efficient and accurate when performing repetitive tasks with defined boundaries, whereas humans may become fatigued when performing such repetitive tasks with negative impacts on accuracy (Schleiger et al., 2023). Human accuracy may also be affected by cognitive biases (Kahneman, 2011), social pressures (Levine, 1999), or stress (FeldmanHall et al., 2015). On the other hand, AI is less effective in novel or ambiguous task settings, and those requiring manual dexterity or the processing of multi-sensory input – all attributes that are strengths of humans (Schleiger et al., 2023).

  One approach to this division of strengths and weaknesses is to continue to push for more and more autonomous AI; training machines to be more human, more capable of the activities humans are capable of, and requiring ever-diminishing human input and intervention. An alternative approach is collaborative human-AI (HAI) teaming. In this approach, rather than developing machines that can more closely emulate humans, the focus is on developing AI that can work with humans on a shared objective to leverage complementary strengths and weaknesses. In doing so, the human and the machine may produce something that neither would have been able to do alone (Paris & Reeson, 2021, Patel et al., 2019; Phillips et al., 2018; Steyvers et al., 2022).

  O'Neill and colleagues (2020) set out key definitional criteria that distinguish collaborative HAI teaming from other forms of human-AI interaction. As a starting point, the team must include at least one autonomous machine agent and one human agent working to achieve a shared objective. Achieving this shared objective must require interdependence of the agents and leverage their complementary capabilities. To these criteria, we add that collaborative HAI teaming must extend beyond a single or one-off interaction and involve the human and machine agent communicating and responding to one another's input and feedback over time. Given these criteria, the benefits of a collaborative HAI teaming approach are most likely to be realised in contexts that are complex, dynamic, or ambiguous, and have task components where humans continue to hold a comparative advantage over machines. One example is collaborative monitoring for anomaly detection and surveillance, which has applications in a range of domains, including radio astronomy, cybersecurity and more. In the specific context of highly complex, multi-element radio telescopes, for example, collaborative monitoring may address the general challenge of anomaly detection in control systems and data streams that are susceptible to external and environmental influences. A collaborative HAI teaming approach has the potential to maximize operational efficiency by navigating a complex ecosystem of anomalies. Machine data cleaning, automated anomaly detection and cross-referencing may be collaboratively combined with human contextual knowledge to identify common causes and maximize the potential to make new scientific discoveries.[1]

Trust is a critical element in appropriate reliance on, and compliance with, any automated or autonomous system (e.g., Hoff & Bashir, 2015; Lee & See, 2004). Trust only becomes more important as such systems grow more complex and the processes by which they come to their outcomes are increasingly opaque. When we have little hope of understanding *how* a system has come to a particular decision or recommendation, it is often our level of trust that mitigates the accompanying uncertainty and guides our choice to use the system (Jacovi et al., 2021). Many models exist accounting for potential antecedents and moderators of trust in automation and AI (A/AI), with most holding reliance as the primary outcome of interest (e.g., Hoff & Bashir, 2015; Kaplan et al., 2021; Lee & See, 2004). Trust in collaborative HAI teams, however, will be necessary to provide a foundation not only for appropriate reliance and compliance, but for the sustained working relationship that characterizes collaborative teaming. In all-human teams, trust fulfils a similar role of mitigating uncertainty and limiting the need for time-consuming control and monitoring efforts (De Jong et al., 2016), as well as providing a basis for effective team communication (Berry, 2011). It is especially relevant in the collaborative contexts requiring interdependence, knowledge sharing, and adaptability that differentiate HAI teams from traditional human-AI interaction (Choi & Cho, 2019; Costa et al., 2018).

---

[1] For more examples of science applications for HAI teams see https://research.csiro.au/cintel

Given the unique requirements of collaboration and the novel pairing of human and machine capabilities in collaborative HAI teams, it is likely that the dynamics of trust formation and outcomes in these applications may diverge from established models of trust in A/AI. In this article we present an overview of current models of interpersonal trust and trust in A/AI with the goal of identifying factors that are likely to be relevant in the context of collaborative HAI teaming. We then draw on these factors to propose a framework for the antecedents, processes, and outcomes of trust in collaborative HAI teaming. In doing so we contribute to the growing science of HAI teaming and identify implications and fruitful future directions for design and research.

**2. Foundations of the trust concept**

Trust is a complex, multi-faceted construct investigated in many contexts, with a multitude of methods, and linked to an array of socially significant outcomes. The scientific literature is replete with conceptions of trust; however, the most established and widely used tend to converge on a set of key features highlighted in Table 1; namely, (a) an expectation or belief that (b) a specific subject will (c) perform future actions with the intention of producing (d) positive outcomes for the trustor in (e) situations characterized by risk and vulnerability (Castaldo et al., 2010).

Table 1. Definitions of trust demonstrating convergence on five key features

| Definition | Features |
|---|---|
| Expectancy held by an individual that the word, promise or written communication of another can be relied upon (Rotter, 1967) | (a) Expectation/belief<br>(b) Specific subject<br>(d) Positive outcomes |
| A state involving confident positive expectations about another's motives with respect to oneself in situations entailing risk (Boon & Holmes, 1991) | (a) Expectation/belief<br>(d) Positive outcomes<br>(c) Risk/vulnerability |
| The willingness of a party to be vulnerable to the actions of another party based on the expectation that the other will perform a particular action important to the trustor, irrespective of the ability to monitor or control the other party (Mayer et al., 1995) | (a) Expectation/belief<br>(b) Specific subject<br>(c) Future actions<br>(d) Positive outcomes<br>(e) Risk/vulnerability |
| The specific expectation that another's actions will be beneficial rather than detrimental and the generalized ability to take for granted … a vast array of features of the social order (Creed & Miles, 1996) | (a) Expectation/belief<br>(b) Specific subject<br>(c) Future actions<br>(d) Positive outcomes |
| Confident positive expectations regarding another's conduct in a context of risk (Lewicki et al., 1998) | (a) Expectation/belief<br>(b) Specific subject<br>(d) Positive outcomes<br>(e) Risk/vulnerability |
| A psychological state comprising the intention to accept vulnerability [to another] based upon positive expectations of the intentions or behavior of another (Rousseau, 1998) | (a) Expectation/belief<br>(b) Specific subject<br>(d) Positive outcomes<br>(e) Risk/vulnerability |
| The attitude that an agent will help achieve an individual's goals in a situation characterized by uncertainty and vulnerability (Lee and See, 2004) | (a) Expectation/belief<br>(b) Specific subject<br>(c) Future actions<br>(d) Positive outcomes<br>(e) Risk/vulnerability |

## *2.1 Models of interpersonal trust*

Mayer, Davis and Schoorman's (1995) model of trust is one of the most widely cited in the interpersonal trust literature and was one of the first models to explicitly delineate trust from trustworthiness (Alarcon et al., 2020; Hoff & Bashir, 2015). Their Integrative Model of Organizational Trust was also one of the earliest models to articulate the role of vulnerability in the trust construct (Baker et al., 2018). In this model, trust is predicted by specific characteristics of both the trustor and the trustee. When considering the trustor, it is readily apparent that some people are simply more willing to extend their trust than others (Alarcon et al., 2018). This trait-based tendency to trust - independent of the target of that trust - is referred to as *trust propensity* by Mayer et al (1995) and elsewhere in the literature as *dispositional trust* (e.g., Hoff & Bashir, 2015).

The primary characteristic of the trustee in this model is their perceived trustworthiness along three dimensions; benevolence, integrity, and ability. Benevolence is an evaluation of the extent to which the trustee has the trustor's best interests at heart. It is based in beliefs around attachment and loyalty between the parties and usually has an affective component (Breuer et al., 2020; Costa et al., 2018).

Integrity reflects the perceived moral standing of the trustee and the degree to which they are believed to be motivated by social norms and principles acceptable to the trustor (Mayer et al., 1995; Schoorman et al., 2007). Trust, however, does not rest on good intentions alone. Ability refers to a target's capacity to execute the behaviors expected of them and is analogous to notions of competence and expertise (Mayer et al., 1995; Schoorman et al., 2007).

Decades of empirical investigation based on this model has provided further insight into the nature of interpersonal trust. Addison (2013) found that the weightings of the three trustworthiness factors in predicting trust vary across cohorts according to differences in interdependence. In a study integrating prior research with a series of qualitative interviews, Breuer et al (2020) identified predictability and transparency as additional dimensions of perceived trustworthiness. Transparency reflects the trustee's clear and open sharing of information, while predictability, as distinct from ability, reflects a consistency or regularity of behavior (Dietz & Den Hartog, 2006). Finally, a meta-analysis of trust antecedents found that although trustworthiness perceptions and trust propensity are correlated, they each have an independent effect on trust formation (Colquitt et al., 2007). This meta-analysis also foreshadowed Breuer et al's (2020) findings of a positive relationship between trust and expressions of relational risk-taking, including disclosure, reliance, and contact-seeking.

*2.2 Models of trust in A/AI*

Many approaches to trust in A/AI draw on the framework established by Mayer et al (1995) and refined in the interpersonal trust literature, in particular its differentiation between the influence of trustor and trustee characteristics in trust development (Schaefer et al., 2020). Various attempts have also been made to map the ability, benevolence, and integrity dimensions of trustworthiness to perceptions of A/AI systems (Alarcon et al., 2020).

One of the earliest and most influential explorations of the role of trust in automated systems is that of Lee and See (2004). This model was developed specifically to characterize the role of trust in facilitating reliance on technology and echoes Mayer et al.'s (1995) tripartite categorization of trust antecedents into individual, contextual, and trustee characteristics. The individual characteristics identified as contributing to trust development are trust propensity and the nature of an individual's previous experience with automated systems. Contextual factors are linked to characteristics of the organization and culture in which the system is embedded, for example social norms and expectations and the attitude of third parties. Finally, trustee characteristics are described exclusively in terms of perceived trustworthiness and linked explicitly to Mayer et al's (1995) dimensions of ability, integrity and benevolence. The authors nonetheless recognize that ability, integrity, and benevolence as articulated by Mayer et al (1995) have uniquely human implications. Benevolence, for example, arguably cannot exist in the absence of intention and to this day (and certainly at the time Lee and See developed their model in 2004) purely technological systems have not attained the capacity for intention (Hatherley, 2020). Consequently, Lee and See recast their dimensions of trustworthiness in terms more applicable to machine agents, that is characteristics of performance, process, and purpose.

Performance is readily identified as the automation corollary of the interpersonal trustworthiness dimension of ability. It is concerned with the current and historical reliability, accuracy and predictability of the system. Process characteristics are concerned with the system's fit to the situation and goals of the operator and is intended to correspond to human integrity, that is, a human trustee's adherence to norms and principle that are acceptable to the trustor. Trustworthiness perceptions based on these process characteristics reflect an evaluation of the nature of the system itself, rather than simply its actions. Finally, purpose is an application of the notion of benevolence to the operation of automated systems. To accommodate a machine's lack of capacity for good intentions, trustworthiness evaluations of the purpose of technology are proposed to be a proxy for perceptions of the designer of the system. Specifically, Lee and See (2004) define this characteristic as reflecting the intention of the designer in developing the system.

In 2015, Hoff and Bashir updated and extended Lee and See's (2004) work with their three-layer model of trust in automation. Once again, the aim of the model is to characterize trust in relation to the specific outcome of reliance. They bring forward the now-familiar tripartite framework of antecedents as individual-, environmental- and automation-related sources of variability in trust. In a slight departure from previous models, they identify the outcome of this variation as different forms of trust rather than different contributions to a multi-dimensional trust construct. The three forms of trust proposed to contribute to reliance on an automated system are dispositional trust, situational trust, and learned trust.

Dispositional trust is entirely founded on an adaptation of Mayer et al's (1995) trust propensity. Dispositional trust in the Hoff and Bashir (2015) model is specific to automation, that is an individual's' tendency to trust automation independent of the context or the specific system. Other individual-level characteristics such as age, gender, culture, and personality traits are proposed to inform this relatively stable propensity to trust automation. Situational trust is an outcome of properties of both the individual and the environment. Environmental antecedents of situational trust include task difficulty, workload and risk level, while individual antecedents are more transient characteristics of the human trustor, including self-confidence, domain expertise, mood, and attentional capacity. Finally, learned trust is formed in response to experience of the performance of the technology. Design features or attributes of the system, such as appearance and communication style, are only believed to be relevant to trust to the extent that they are perceived to affect system performance.

Other significant models of trust in A/AI diverge further from theories of interpersonal trust, retaining the tripartite structure of antecedents, but breaking with efforts to fit technological factors into Mayer et al's (1995) human dimensions of perceived trustworthiness. Hancock et al's (2011) literature review and meta-analysis identifies three categories of antecedents of trust specifically in robots. Consistent with the tripartite framework, they are labelled as human-related, robot-related and environmental factors. Hancock and colleagues further split each category of antecedents into two discrete aspects. Human-related antecedents are divided between those based in ability, for example expertise, competence and situational awareness, and those based in individual characteristics, including demographics, self-confidence, and trust propensity. Robot-related antecedents are similarly distributed between those grounded in performance, and those describing attributes of the robot, including adaptability and anthropomorphism. Environmental antecedents may relate either to team factors, such as communication, or task factors like task complexity and physical environment. The meta-analysis indicated minimal influence of human and environmental factors on development of trust in robots, identifying robot-related factors as the primary driver of trust formation.

Kaplan et al (2021) recently used the same tripartite categories to investigate trust in AI, believing that its complexity and ubiquity may contribute to trust dynamics that differ from those of either more static and predictable automation (Schaefer et al., 2016), or robots that are confined to specific roles and environments (Hancock et al., 2011, 2021). Indeed, this meta-analysis found that both human-related and environmental antecedents significantly predict trust in AI. Human-related factors with a particularly strong influence on trust formation included competence, expertise, culture, and gender, while highly relevant environmental factors included length of the team relationship and risk level. Unsurprisingly, factors of the technology were also found to play a large role in trust development, in particular performance, reliability, human-centred language, anthropomorphism, and perceptions of transparent behavior by the system.

For both robot and AI trust referents, teaming capability was a predictor of trust, and the appropriateness of cues and feedback from the system to the human were also highly relevant (Hancock et al., 2011; Kaplan et al., 2021). This suggests that elements of the technology and environment that imply or support collaborative working are likely to have a positive influence on human trust (Schaefer et al., 2020). Consequently, just as we see a shift in antecedents between automation, robots, and AI, we may expect to see a novel constellation of factors predict trust in expressly collaborative forms of AI.

## 3. A process framework of trust in collaborative HAI teams (CHAI-T)

A recent report from the National Academies of Sciences, Engineering, and Medicine (Committee on Human-System Integration Research Topics for the 711th Human Performance Wing of the Air Force Research Laboratory et al., 2022) on the state-of-the-art in HAI teaming identified the need to develop models of trust that 1) account for the specificity of task contexts and goals, 2) integrate processes of team interaction, and 3) capture how trust evolves over time. Certainly, no single set of factors and processes can be expected to capture the formation and maintenance of trust across the range of potential applications of collaborative AI. To meet the needs of this proliferation, a model of trust must have the capacity to respond to the specific contextual conditions of the trustor, the trustee and the collaborative task environment. Existing models of trust in A/AI tend to put forward a single set of antecedent factors cast as being agnostic to the particular type of technology that is the target of trust. Yet meta-analyses show that factors influencing trust formation differ between targets (Hancock et al., 2011; Kaplan et al., 2021). These differences may be qualitative, that is factors that are relevant to one form of technology (for example, robots) do not similarly influence trust in other systems (for example, disembodied AI algorithms). The differences may also be quantitative, with the factor being relevant to multiple systems, but carrying different weight across them (for example, anthropomorphism may contribute to trust in both robots and disembodied algorithms but play a greater role in trust in robots). Factors shown to be relevant to human-teaming, such as communication capability and team tenure, are likely to take on more prominence in a collaborative HAI interaction (Wang et al., 2016; Wynne & Lyons, 2018).

Traditional models of trust in A/AI also do not tend to take account of how the actual process of collaboration in HAI teams may influence patterns of trust development. There is extensive scientific literature, however, on the relationship between trust and team interaction in all-human teams. Features of interaction shown to have an effect on trust in this context include communication frequency (Becerra & Gupta, 2003), information sharing (Fulmer & Gelfand, 2012), and shared awareness of team members, tasks, and situations (Muir, 1994). These insights from the psychological and management sciences have the potential to inform approaches to trust in teams comprising both human and machine agents.

Although many models recognize the role of feedback and recalibration, in practice there has been little exploration of the temporal dynamics of trust formation and maintenance, especially in the context of longitudinal teaming (de Visser et al., 2020). Mayer et al's (1995) early model of trust between humans includes explicit feedback loops between the outcomes of trust and trustworthiness perceptions, yet empirical testing of pathways in this model mostly neglect these feedback loops (Costa et al., 2018). We see similar disjunctions between theory and empirical investigation for models of trust in A/AI (de Visser et al., 2020). Yet research specifically exploring temporal trajectories of trust development towards both human and A/AI targets clearly indicates that this process is highly dynamic.

Coovert et al (2017) find, for example, that early performance has a stronger impact on trust in human co-workers than later evaluations of performance. They also find that the extent of this impact differs across individual trustors. Similarly for trust in A/AI, individual differences among humans have a strong influence on trust early in an interaction when the operator has little direct knowledge of the system (Alarcon et al., 2020). There is also evidence to suggest that trust development trajectories differ systematically between targets. Trust in intelligent robots tends to start low and increase over time, while the opposite pattern has been observed for disembodied AI (Lockey et al., 2021). Understanding the specific trajectories of trust formation and maintenance across a range of collaborative AI applications will be critical to developing long-term teaming capacity and ensuring appropriate trust calibration.

In the next section we draw on process models of human teaming to propose a framework of human trust in collaborative HAI applications that incorporates these requirements for context specificity, interaction processes, and temporal dynamics.

### 3.1 Input-Mediator-Output Models of Teaming

A process model in the human teaming context is one in which team inputs are converted to goal-based outcomes through cognitive, verbal and behavioral processes of the team (Ilgen et al., 2005). A frequently drawn on process model framework is the IMOI (Inputs-Mediators-Output-Inputs) model (see Figure 1). Inputs are characteristics of the members and context of a team, i.e., the antecedent categories of human, technology and contextual factors. Mediators include both team processes and dynamic emergent states, like trust, that can both shape, and be the product of, team experiences. Outputs most commonly reflect team effectiveness in some form, and the specific outcome of interest may depend on where the team is within its life cycle. In early stages, for example, the key outcome may be collaboration, whereas as the team gains tenure and becomes more established, task performance metrics become more relevant. Finally, the IMOI model circles back to inputs, where mediators (like trust) and outputs serve as inputs in a new cycle (Ilgen et al., 2005; Marks et al., 2001).

Figure 1. Inputs-Mediators-Outputs-Inputs process model of teaming

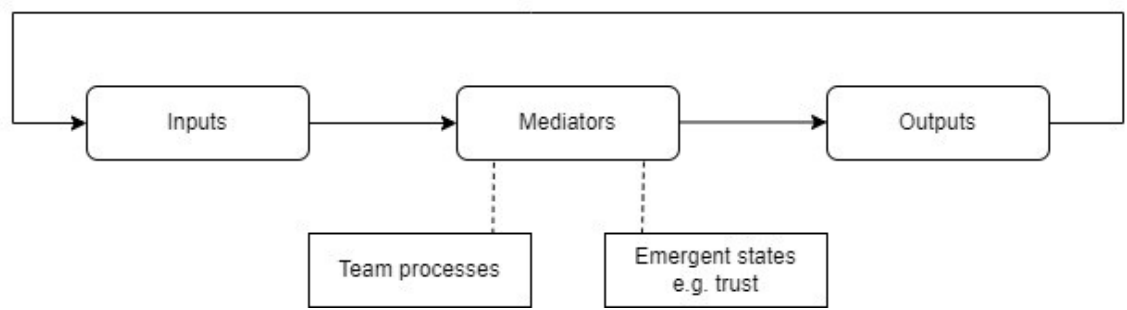

IMOI models have been described as offering "a way to unpack the 'black box' representing the processes and emergent states through which teams transform inputs into valued outputs" (O'Neill et al., 2023, p. 4). With such allusions to black boxes, it is perhaps not surprising to see increasing calls for these models to be integrated into investigations of teaming between humans and AI agents. In the introduction to a recent special issue on advances in human-autonomy teamwork, O'Neill and colleagues (2023) promoted IMOI models as a rigorous theoretical approach to delineating mediating constructs, such as trust, from other outcomes of interest in teams (e.g. performance), and investigating changes in trust over the life cycle of a team. Consistent with these recommendations, CHAI-T is not a one-size-fits-all model of trust in collaborative HAI teams, but rather, a than framework of components that may come together in different ways and different contexts to product and sustain trust.

*3.2 Context specificity in the CHAI-T framework*

The inputs available to human teams vary; for example, by the number of team members, the length of time in the team, the training and expertise of individual members, or the time constraints associated with the team task. The inputs available to the development of trust in collaborative HAI teams may vary similarly. The factors that contribute to trust in an embodied robot, for example, are likely to be different to the factors relevant to the formation of trust in an algorithm supporting collaborative digital data curation. Each of these applications requires its own 'recipe' for trust, and each of these recipes may work best with particular ingredients. Indeed, with more than 470 distinct antecedents to trust in AI identified in the literature, the pantry of ingredients is so large that models may become bloated or lack ecological validity if they attempt to include all possible antecedent factors (Saßmannshausen et al., 2022). To develop recipes for trust in specific collaborative AI applications, researchers will require a comprehensive understanding of the sociotechnical context in which the AI system is intended to operate, including characteristics of the users, the task, and the wider organisational environment.

Within IMOI models, outputs are also context specific. Effective operation of collaborative HAI teams requires that we understand the desired outcomes for a particular team at a given stage in its life cycle. Trust in collaborative AI must always serve a purpose. The positioning of trust as a mediator in an IMOI model clarifies the role of trust as a means to an end - whether that be increased reliance, adoption, or performance – not an end in itself (O'Neill et al., 2023). In the early stages of a team, the desired outcome of trust may be the facilitation of behaviors enabling collaboration, for example allocation of control authority to the system, refraining from unnecessary monitoring, and willingness to rely on the system (Breur et al, 2020; Schaefer et al., 2020). Once sufficient integration of the human and AI system has occurred, trust may contribute more directly to task and team performance. It is important to note, however, that the optimal level of trust for achieving team objectives is not equivalent to the maximal level of trust. Trust in a machine system is considered to have reached its optimal level and, consequently, best facilitate performance outcomes, when the human trustor's expectations are aligned with the capabilities of system (Jacovi et al., 2021; Schaefer et al., 2020). When optimal levels of trust are exceeded, that is, when humans expect more of a system than its capabilities warrant, the resulting overreliance and lack of monitoring can have catastrophic outcomes. A failure to reach optimal levels of trust, on the other hand, can result in disuse of technological systems, representing a waste of time and resources (Parasuraman, 1997).

To take account of the context specificity of trust outcomes in collaborative HAI teaming, we again recommend researchers begin with comprehensive profiling of the team and task environment. By identifying relevant team metrics across the lifespan, it becomes possible to experimentally quantify the degree of trust required to achieve the desired outputs.

*3.3 Interaction processes*

Mediators play a key role in IMOI models of teaming, and especially in the CHAI-T process framework. Our focal construct, trust, is a mediator between team inputs and outputs in this framework. Just as significant, however, are the processes of interaction or team processes, that also influence how individual, technological and environmental inputs come together to produce trust and contingent team outputs.

Team processes are the interdependent activities that convert inputs to outputs, often indirectly via mediating constructs like trust. They are fundamental to facilitating collaboration and enhancing team effectiveness (Kozlowski & Chao, 2018). If we conceive of trust in a collaborative AI application requiring a certain recipe and antecedent factors as ingredients, then the team processes of the framework are, quite straightforwardly, the processes by which those ingredients are prepared and transformed into a particular dish. In the proposed framework of trust in collaborative AI, these processes are influenced by inputs, the trust construct, and the model outputs. The processes in turn work upon trust and potentially the outputs directly (see Figure 2).

Figure 2. The CHAI-T process framework

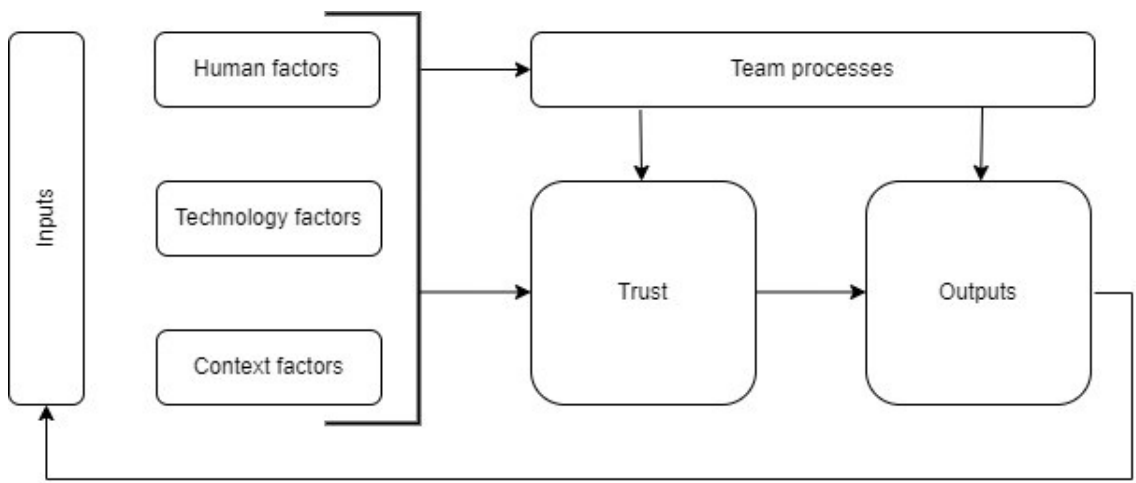

A wide range of team processes have been investigated in the human teaming literature, which includes taxonomies and measures of specific processes. Marks and colleagues (2001) offer an empirically supported taxonomy of ten team processes (see

Table 2), with the recommendation being to include in any given IMOI model the processes most relevant to the specific teams and tasks under study (LePine et al., 2008). Many of these processes have intuitively plausible parallels to the nature of collaborative interactions between humans and AI. For example, formulation and planning is analogous to training the human and programming the machine counterparts in the collaborative HAI team, including setting initial expectations, learning the parameters of the system, and identifying team goals. Monitoring progress towards goals involves tracking progress, interpreting information in terms of what still needs to be done, and sharing that information with team members. System transparency, communication, and situational awareness may contribute to this process, and in doing so may influence trust (Atkinson et al., 2014; Chen et al., 2018). In team monitoring processes, team members observe the actions of their teammates – in this case the AI system – and watch for errors or performance discrepancies. This is another means by which a system's capacity to communicate uncertainty may contribute to trust formation and is a key location for trust calibration in the framework. Finally, coordination is the process of orchestrating the sequence and timing of interdependent actions. It involves mutual information exchange and adjustment of action and is where features of collaborative workflows between human and AI may be expected to interact with trust formation and maintenance.

Table 2. Taxonomy of team processes in human teams (Marks et al., 2001)

| Process | Description |
|---|---|
| Mission analysis | Identification and evaluation of team tasks, challenges, environmental conditions, and resources available for performing the team's work |
| Goal specification | Activities centred on identification and prioritisation of team goals |
| Strategy formulation and planning | Developing courses of actions and contingency plans; making adjustments to plans in light of changes or expected changes in a team's environment |
| Monitoring progress towards goals | Paying attention to, interpreting, and communicating information necessary for the team to gauge its progress toward its goals |
| Systems monitoring | Tracking team resources and factors in the team environment to ensure the team has what it needs to accomplish its goals and objectives |
| Team monitoring and back-up behavior | Team members assisting other members in the performance of their tasks |
| Coordination | Synchronising or aligning the activities of team members with respect to their sequence and timing. |
| Conflict management | How conflict is proactively and reactively dealt with |
| Motivating and confidence building | Activities that develop and maintain motivation and confidence in relation to accomplishing team goals and objectives |
| Affect management | Activities that foster emotional balance, togetherness, and effective coping |

Despite their apparent compatibility with collaborative HAI teams, it cannot be assumed the interaction processes of human teaming are able to be meaningfully transposed to a HAI teaming context. Experimental studies should also be undertaken to test the effectiveness of instantiating adaptations of human team processes in collaborative HAI teaming.

### *3.4 Temporal dynamics of the CHAI-T framework*

Collaborative HAI teaming is further distinguished from other forms of human-AI interaction by its particular temporal dynamics. This approach to human-AI interaction is not intended as a single or one-off interaction, nor is it static. Collaborative HAI teaming involves a sustained interaction in which the human and the AI system are communicating and responding to one another's input and feedback (Schleiger et al., 2023). Using IMOI models of teaming it is possible to capture the trajectory of trust development, maintenance, and influence over time via the incorporation of recurring performance episodes.

Performance episodes may be defined as a "distinguishable period of time over which performance accrues and feedback is available … Episodes are most easily identified by goals and goal accomplishment periods, and the conclusion of one period normally marks the initiation of another" (Marks et al., 2001, p. 359). A HAI team engaged in collaborative monitoring, for example, may track an anomalous system reading and alert a human team member if it is unable to identify its nature or source. The human may then act on the alert, provide feedback to the AI system, or take another form of action. This resolution would represent the end of a performance episode for the team in question. The nature of performance episodes will vary by team and task type, reinforcing the need for models of trust in collaborative HAI teams to be sensitive to context. In some contexts, episodes may be easily delineated; in others the distinction between one episode and the next is less clear cut. Once again, detailed profiling of the task environment, with a particular attention paid to workflows, will be required to identify performance episodes for a given application.

In a multi-phasic IMOI model, the outputs from one episode become one of the inputs for the next cycle (see Figure 2). In this way, levels of trust and the outputs that flow from it influence the input, processes, and outputs of subsequent performance episodes in a cyclical pattern that allows for dynamic changes in trust and team outcomes over time (Demir et al., 2021; Larson et al., 2020). For example, an AI system's ability to explain its behavior or recommendations (a technology-based input to the IMOI model) may be critical to develop human trust during the early stages of teaming but play a smaller role in sustaining that trust over the longer term. Similarly, close monitoring of AI system performance (a teaming process) may contribute to building appropriate trust in the early phases but become less necessary over time. The CHAI-T framework permits identification of factors related to the interaction between human and machine that increase or decrease levels of trust, and in doing so, provides a framework for active trust management.

**4.0 Application of the CHAI-T framework**

Collaborative HAI teaming is the next frontier of interaction between humans and machines. To fully realise the productive potential of complementary human and machine strengths, we require a science of HAI teaming that provides a theoretical framework to incorporate the wide range of dynamic factors in teaming contexts (McNeese et al., 2023). Just as productivity in all-human teams benefits from systematization of knowledge in process frameworks, research into collaborative HAI teams will have its greatest impact when unified in an overarching theoretical approach. Early calls for the exploitation of IMOI models in the collaborative HAI teaming domain have identified a significant gap in examinations of the constructs and processes that mediate the relationships between team inputs and outputs. Trust is one such key mediator in teaming. Effective conversion of team inputs into desired outputs is often contingent on developing an optimal degree of trust (Schaefer et al., 2021). In this paper we have presented an IMOI framework of trust in collaborative AI which takes into account the context specificity, team processes, and temporal dynamics that characterize HAI teaming.

To use the CHAI-T framework to specify a model of trust in a particular collaborative AI application we have recommended that researchers begin by developing a comprehensive profile of the HAI team and task. This should include, but not be limited to, identification of team goals, tasks and their risk profile, determination of performance episodes, the types of users expected to interact with the system, desired team outputs and relevant team processes. There are numerous ways to collect the data to inform such profiles and the most appropriate method will be informed by the nature of the task and available human participants. For example, one approach may be to undertake a dedicated survey of potential human users. Alternatively, relevant questions may be incorporated into existing data collection exercises, such as user studies. Where domains have been subject to significant investigation, detailed profiles may be built from reviews of the literature. In novel domains, or those without a readily accessible cohort of users, profiles may be informed by expert small group discussions.

Once an application profile is completed, researchers can cross-reference key characteristics of the application with empirically verified trust antecedents in the scientific literature. A current challenge to this approach, however, lies in navigating this large and unstructured body of knowledge. Recent meta-analyses, such as that from Saßmannshausen and colleagues (2022) provide a starting point with an index of hundreds of antecedent factors categorized into human, technology, and environmental characteristics. We are currently developing a searchable database of antecedent factors that we plan to make available to researchers to more efficiently facilitate this cross-referencing process.

The trust antecedents identified as being relevant to the collaborative HAI team under investigation may then be entered into a model of trust and subject to hypothesis testing. This may include empirically verifying the contribution of those input factors to trust in collaborative AI, and may extend to testing the role of various team processes in trust development and maintenance. Once specified and validated a model can be used to test the optimal level of trust for desired outcomes and the features of team and task design necessary to achieve and maintain that trust (O'Neill et al., 2023).

The incorporation of team processes and temporal dynamics into the CHAI-T framework permits the integration of a range of empirical findings into a single unifying framework of human trust in emerging AI interaction paradigms. The power of process models to facilitate such theoretical integration is already acknowledged, with calls for the IMOI framework to be adopted as a common general model of HAI teaming collaboration (O'Neill et al., 2023). As an example, it is widely recognized that transparency plays a role in trust formation and maintenance, but that the nature of the relationship between them is not straightforward (Baker et al., 2018; Lockey et al., 2021). While transparency may foster trust under some circumstances, at other times greater transparency may interact with human cognitive capacity to ultimately reduce trust (Dikmen & Burns, 2022). Our proposed trust framework allows for transparency to operate via different processes and with different effects at different stages in the life cycle of the team. The contribution of transparency to team monitoring processes may, for instance, increase trust with beneficial outcomes for collaboration in the early 'team-building' phases of the human-AI relationship. At later stages when team performance becomes a higher priority outcome, transparency information may be of lower value to both trust and team functioning. The framework also accommodates advances in the shared understanding and situational awareness of AI systems. Chen et al's (2020) model of Dynamic Situation-Awareness-based Agent Transparency (Dynamic SAT) describes the information machine agents must convey about their decision-making processes to facilitate the shared understanding that underpins effective teamwork. They identify the need to convey task information such as task procedures, likely scenarios and environmental constraints, which, in our framework, would influence and interact with trust via the team process of monitoring progress towards goals. Similarly, the team-related information they identify, such as role interdependencies, interaction patterns, and relative abilities, may interact with trust via other team processes including formulation and planning, team monitoring, and coordination.

A key contribution of the CHAI-T framework to efficient HAI teaming is its facilitation of active trust management. The field currently lacks well-developed methods for aligning trust-based expectations with actual AI system performance and identifying when optimal levels of trust have been achieved (Jacovi et al., 2021; Schaefer et al., 2020). The CHAI-T process framework advances these efforts in a number of ways. The first is by capturing changes in trust levels over time. In doing so, it permits the identification of factors in the interaction between human and machine that increase or decrease levels of trust. Secondly, the framework explicitly identifies team processes expected to facilitate this calibration of expectation and performance, including the monitoring of progress towards goals, monitoring the performance of team members (including the AI system), and communication of current status. In this way, the framework can be drawn on to inform the design of system capabilities, collaborative workflows, and training of human team members in a way that supports trust calibration. For example, specifying the team processes that play a role in trust formation and maintenance can guide development of collaborative AI capabilities, such as communication, situational awareness, and transparency. Specifying the characteristics of the human user that influence trust development and maintenance may also inform skills training and interventions (Committee on Human-System Integration

Research Topics for the 711th Human Performance Wing of the Air Force Research Laboratory et al., 2022).

**5.0 Future directions for research**

The CHAI-T framework integrates top-down, theory-informed approaches with bottom-up, data-driven validation to facilitate the specification of verifiable models of trust development for individual applications. The allowance for context-specificity that is a strength of the framework, also requires that researchers adopt effective methods for identifying the antecedent factors (or ingredients) that will build a recipe for trust in their target application. We are currently testing approaches to specifying trust models for collaborative AI applications across a number of use cases in the science domain. The incorporation of team processes into the framework also presents a number of key directions for future research. As noted in section 3.3, human teaming processes may not be meaningfully applicable in a HAI context. Further investigation is required to determine which teaming processes are relevant to HAI teaming in a general sense, as well as means of determining which processes are likely to have an influence on trust and performance in a particular collaborative HAI team.

The CHAI-T framework offers a rigorous theoretical approach to specifying models of trust in emergent forms of human-AI interaction. However, effective validation of the framework and the specific trust models it informs will require equally rigorous methodological approaches. Research methods are required that can validly and reliably measure a latent variable like trust. Numerous methods exist for quantifying trust, each with their own limitations (French et al., 2018). Self-report measures such as surveys can be time-consuming and may not map consistently to relevant decisions and behavior (Baker et al., 2018; Miller, 2022). Observed behavior may be influenced by factors other than trust such as fatigue or cognitive load, while physiological measures remain imprecise in what they can convey about underlying cognitive processes (Schaefer et al., 2020). We need psychometrically validated self-report scales and multi-method approaches to trust measurement that meet the practical constraints of the team and task environment.

Efficiently and accurately measuring temporal phenomena poses another key challenge in devising research approaches to testing a recurring phase model of trust. Common questions across both the all-human and HAI teaming domains include the appropriate rate of sampling to capture the processes of interest, and the identification of data collection methods that are "cheap, frequent, and stealthy" (Kozlowski, 2015, p. 280). Potential avenues of exploration include experience sampling methods (Van Fossen et al., 2021), use of behavioral sensors or video observations (Grand et al., 2013), and computational modelling (Kozlowski, 2015).

**6.0 Conclusion**

Trust has been shown to be integral to human interaction with automation and AI. Models of trust in A/AI predominantly draw on a tripartite structure of antecedent categories inherited from investigations of trust among humans; human or trustor characteristics, technology or trustee characteristics, and characteristics of the environment or context.

Collaborative HAI teaming represents a new approach to human-AI interaction that complements the advantages of traditional AI with the unique strengths of human capabilities and values in a process of augmentation rather than automation. Human trust in collaboration with AI may be more likely to draw on the dynamics of human-human trust to a greater extent than human trust in other technologies. The CHAI-T process framework proposed in this paper is intended to respond to the particular requirements of collaborative HAI teams. It is primarily distinguished from earlier models of trust in A/AI by its incorporation of team processes and recurring performance phases, across which trust and its outputs are influenced by previous team experiences and outcomes. These features enable active management of trust calibration in collaborative HAI teams, with practical implications for the design of team processes, system capabilities, and human training. In doing so, the CHAI-T process framework offers a single unifying approach to modelling the optimal level of trust required for safe and effective human-AI collaboration across a range of applications.